\documentclass[aps,prd,onecolumn,nofootinbib,showkeys,amssymb,11pt]{revtex4}
\usepackage[a4paper,left=20mm,right=20mm,top=25mm,bottom=25mm]{geometry}
\usepackage{amsmath}
\usepackage{amsfonts}
\usepackage{amssymb}
\usepackage[utf8]{inputenc}
\usepackage[T1]{fontenc}
\usepackage[colorlinks=true]{hyperref}
\usepackage[dvipsnames]{xcolor}
\usepackage{graphicx}
\usepackage[normalem]{ulem}
\usepackage{float}
\usepackage{amssymb}
\usepackage{amsfonts,color}

\begin{document}

\title{Warm Higgs-Starobinsky inflation}

\author{Daris Samart}
\email{darisa@kku.ac.th}
\affiliation{Khon Kaen Particle Physics and Cosmology Theory Group (KKPaCT), Department of Physics, Faculty of Science, Khon Kaen University,123 Mitraphap Rd., Khon Kaen, 40002, Thailand}

\author{Patinya Ma-adlerd}
\email{m.patinya@kkumail.com}
\affiliation{Khon Kaen Particle Physics and Cosmology Theory Group (KKPaCT), Department of Physics, Faculty of Science, Khon Kaen University,123 Mitraphap Rd., Khon Kaen, 40002, Thailand}

\author{Phongpichit Channuie}
\email{channuie@gmail.com}
\affiliation{College of Graduate Studies, Walailak University, Thasala, Nakhon Si Thammarat, 80160, Thailand}
\affiliation{School of Science, Walailak University, Thasala, Nakhon Si Thammarat, 80160, Thailand}
\affiliation{Research Group in Applied, Computational and Theoretical Science (ACTS), Walailak University, Thasala, Nakhon Si Thammarat, 80160, Thailand}
\affiliation{Research Center for Fundamental Physics (RCFP), Walailak University, Thasala, Nakhon Si Thammarat, 80160, Thailand}

\date{\today}

\vskip 1pc
\begin{abstract}
In this work, we investigate the Higgs-Starobinsky (HS) model in the context of warm inflation scenario. The dissipative parameter as a linear form of temperature of warm inflation is considered with strong and weak regimes. We study the HS model in the Einstein frame with the slow-roll inflation framework. We compute the inflationary observables and then compare with the Plank 2018 data. With the sizeable number of e-folds and proper choices of parameters, we discover that the predictions of warm HS model present in this work are in very good agreement with the latest Planck 2018 results. More importantly, the parameters of the HS model are also constrained by using the data in order to make warm HS inflation successful. 
\end{abstract}

\date{today}

\maketitle

\section{introduction}

Despite the fact that the standard model of cosmology, a.k.a. the Big Bang model, provides a comprehensive explanation for a broad range of observed phenomena including the anisotropy of the cosmic microwave background (CMB) consisting of the small temperature fluctuations in the blackbody radiation left over from the Big Bang and a mechanism for generating the primordial energy density perturbations that are the seed for late
time large structure structure. However, there  are  some  observations  in which  the  traditional Big  Bang  model  fails  to  explain. These cosmological problems are linked to the primordial universe. More concretely, the observed flatness, homogeneity, and the lack of relic monopoles posed severe problems in the standard Big Bang cosmology. In order to solve such fundamental problems, an inflationary scenario \cite{Starobinsky:1980te,Sato:1980yn,Guth:1980zm,Linde:1981mu,Albrecht:1982wi} is a well-established paradigm describing an early universe and posts an indispensable ingredient of modern cosmology.

In the standard picture, accelerated expansion quickly erases all traces of any pre-inflationary matter or radiation density resulting the universe in the vacuum state. We explain the transition from inflation to the “hot Big Bang” state by requiring the nucleosynthesis and using the physics of recombination leading to the descriptions of the CMB temperature anisotropies we observed today. To this end, we need the interactions between the inflaton with other fields resulting the (partial) decay of the inflaton into
ordinary matter and radiation. However, inflaton decay can only play a significant role at the end of the slow-roll regime, leading to the standard “(p)reheating” paradigm, see e.g. \cite{Linde:2005ht,Albrecht:1982mp,Abbott:1982hn}. In standard cold inflation, any preexisting radiation is stretched and dispersed during a very short cosmic phase and no new radiation is produced. However, one can imagine an alternative scenario where dissipative effects and associated particle production can sustain a thermal bath concurrently with the
accelerated expansion of the Universe during inflation. This alternative perspective was known as warm inflationary paradigm. 
The original proponent of warm inflation was proposed by Arjun Berera and his colleagues \cite{Berera:1995ie,Berera:1995wh}. As mentioned in Ref.\cite{Berera:1996fm}, this alternative counterpart is proposed in which the radiation energy density smoothly decreases all during an inflation-like stage and with no discontinuity enters the subsequent radiation-dominated stage.

Beside, the Starobinsky $R^2$ cosmic inflation model \cite{Starobinsky:1980te} and the non-minimal coupling Higgs inflation \cite{Bezrukov:2007ep} are greatly received attention over decades. In particular, Those two models are very successful to explain the mechanisms of the inflationary universe and nicely fit with the observational data.  However, these two models suffer from some fundamental problems per se. On the one hand, Higgs inflation encounters to the unitary problem if we consider single scalar Higgs field as the inflaton only and the Higgs field needs to large at the beginning of the inflationary phase \cite{Burgess:2010zq}. On the other hand, the origin or mechanism to generate the $R^2$ term in the Starobinsky model is still unclear. Fortunately, an attempt to combine and fulfill the Higgs with Starobinksy is successfully done by many authors of Refs. \cite{Salvio:2015kka,Calmet:2016fsr}. This leads to a so called Higgs-Starobinsky (HS) inflation model. The main idea of the HS model is that the Higgs field does couple to the graviton (Ricci scalar) at large coupling and this leads to the $R^2$ emerging from the quantum correction between Higgs and graviton at least at one-loop level. As a result, this model does not suffer from all mentioned problems of the Higgs and Starobinsky inflationary models. Salient features of the HS are that there is no physics beyond standard model of particle physics and the higher curvature term $R^2$ of the Starobinsky inflation is automatically generated by the quantum correction effect. In addition, the unitarity problem of the original Higgs inflation is solved. The HS inflation has been used to study in various aspects  \cite{Wang:2017fuy,Ema:2017rqn,Pi:2017gih,He:2018gyf,Gundhi:2018wyz,Antoniadis:2018ywb,Enckell:2018uic,Samart:2018mow,Ema:2019fdd,Ema:2020zvg,Markkanen:2018bfx,Ghilencea:2018rqg}. However, a study of the HS model in warm inflationary universe has not been reported yet and it is worth investigating it in this work.

The structure of the present work is organized as follows. In Section \ref{formalism}, we set up the (warm) HS inflationary model and study it in the Einstein frame. We then derive the relevant cosmological observables in the warm inflation scenario. In Sec.\ref{compare-data} we compare the theoretical results in the warm HS inflation with the Planck 2018 data. Finally, We conclude our findings in the last section. 

\section{Model Set-up}
\label{formalism}
\subsection{The HS action}
The gravitational action of the HS model with non-minimal coupling to the Ricci scalar and the self-interacting Higgs field is given by
\begin{eqnarray}
S_J = \int d^4x\sqrt{-g}\,\left[ -\frac{1}{2}\,M_p^2\, R - \frac{1}{2}\,\xi\,h^2\,R
+ \frac{1}{2}\,g^{\mu\nu}\partial_\mu h\,\partial_\nu h - \frac{\lambda}{4}\,h^4  - \alpha\,R^2 \,\right],
\label{H-J}
\end{eqnarray}
where the subscript $S_J$ stands for the action in Jordan frame and $M_p^2 = 1/8\pi G$, $\xi$ and $\alpha$ are Planck mass, non-minimal Higgs and $R^2$ Starobinsky term coupling constants, respectively. While the $h$ field is the Higgs scalar field with the standard Higgs potential the self-interacting coupling constant $\lambda$. In the HS model, the large coupling of the Higgs and graviton plays the role as the trigger of the Starobinsky inflation term $R^2$ from the quantum correction \cite{Calmet:2016fsr,Salvio:2015kka}. According to the RG analysis of the HS model at the one-loop level \cite{Markkanen:2018bfx,Ghilencea:2018rqg}, it was shown that the coupling of the $R^2$ term, $\alpha$ is proportional to $\alpha(h) \propto (\xi +1/6)^2\ln(h/\mu)$ where the renormalization scale is set at the Planck mass i.e., $\mu \approx M_p$ and the Higgs field ($h$) is a sub-Planckian field as $h \ll M_p$\,. This is the main mechanism behind the generation of the Starobinsky $R^2$ inflation in the HS model. At the large values of non-minimal coupling $\xi$ and the inflaton (scalaron, $\phi$ see below) and in the slow-roll regime during inflation, we can drop kinetic term of the Higgs field. Then the HS gravity action is given by \cite{Pi:2017gih,He:2018gyf,Samart:2018mow},
\begin{eqnarray}
S_J = \int d^4x\sqrt{-g}\,\left[ -\frac{1}{2}\,M_p^2\, R - \frac{1}{2}\,\xi\,h^2\,R
- \alpha\,R^2 - \frac{\lambda}{4}\,h^4  \,\right].
\label{H-J}
\end{eqnarray}
We can eliminate the non-minimal Higgs coupling term, $\xi\,\sigma^2\,R$\, via the equation of motion of $h$ field. The Euler-Lagrange equation of the Higgs field, $h$ is therefore written by
\begin{eqnarray}
\frac{1}{2}\,\xi\,h^2\,R + \frac{\lambda}{4}\,h^4 = 0\,\quad \Longrightarrow\quad h^2 = -\,6\,\xi\,R/\lambda\,.
\end{eqnarray}
Substituting the Higgs field in above equation. One finds
\begin{eqnarray}
S_J &=& \int d^4x\sqrt{-g}\,M_p^2\left[ -\frac{1}{2}\, R - \frac{1}{12\,M^2} \,R^2  \right],
\label{Star-J}
\\
M^2 &=& \frac{M_p^2}{12\left(\alpha + 3\,\xi^2/(2\,\lambda)\right)}\,. \label{M2}
\end{eqnarray}
The above action is a standard form of the Starobinsky inflation action. We will see in the latter that the scalaron mass, $M_\alpha$ of the pure Starobinsky inflaton field (for $\xi = 0 =\lambda$) is given by
\begin{eqnarray}
M_\alpha^2 = \frac{M_p^2}{12\,\alpha}\,,
\end{eqnarray} 
whereas the scalaron mass of the HS gravity is modified by \cite{He:2018gyf,Samart:2018mow}
\begin{eqnarray}
M^2 = \frac{M_\alpha^2}{1+18\,(\xi^2/\lambda)\,M_\alpha^2/M_p^2}\,. \label{Mde}
\end{eqnarray}
According to the observational constraints of the amplitudes of the curvature perturbation, one finds $M \approx 1.3\times 10^{-5}M_p$ \cite{Faulkner:2006ub}. By using the fixing $M$ parameter, we obtain the relation between three parameters $\xi$, $\alpha$ and $\lambda$ and we will employ action in Eq.(\ref{Star-J}) to work out relevant inflation parameters and fix the parameters from the HS model with the observational data in the next section.

It is very convenient to study the inflation dynamics in the Einstein frame which can be obtained via the conformal transformation. According to the HS action Eq. (\ref{H-J}) in the Jordan frame, we can impose the conformal factor as
\begin{eqnarray}
\Omega^2 = \frac{2}{M_p^2}\,\frac{\partial}{\partial R}\left( \frac12\,M_p^2\,R + \frac{M_p^2}{12\,M^2}\,R^2\right) = 1 + \frac{R}{3\,M^2} \,,
\end{eqnarray}
where the definition of the effective mass $M$ is given in Eq.(\ref{Mde}). The conformal factor, $\Omega^2$ plays important role on transformation of the gravitational action from Jordan frame to Einstein frame. The relation between metric tensors of Jordan and Einstein frames reads,
\begin{eqnarray}
g_{\mu\nu} = \Omega^2\,\widetilde{g}_{\mu\nu}\,.
\end{eqnarray}
We would like to mention that all quantities with ``\,\,$\widetilde{~}$\,\," are represented quantities in the Einstein frame. The Ricci scalar in Jordan frame is written in terms of quantities in Einstein frame as
\begin{eqnarray}
R = \Omega^2\left(\widetilde{R} + 3\,\widetilde{g}^{\mu\nu}\partial_\mu\partial_\nu \ln\Omega^2
- \frac{3}{2}\,\widetilde{g}^{\mu\nu}\partial_\mu\ln\Omega^2\,\partial_\nu\ln\Omega^2 \right).
\end{eqnarray}
More importantly, the scalaron field, $\phi$ in the HS model is introduced via
\begin{eqnarray}
\phi = M_p\sqrt{\frac32}\,\ln \Omega^2 \,.
\end{eqnarray}
\allowdisplaybreaks
\begin{eqnarray}
S_E = \int d^4x\,\sqrt{-\widetilde g}\,\left[ -\frac12\,M_p^2\,\widetilde{R} + \frac12\,\widetilde{g}^{\mu\nu}\,\partial_\mu\phi\,\partial_\nu\phi - V(\phi) \right],
\label{action-E}
\end{eqnarray}
Using the definition of the scalaron field, one can write the effective potential of the scalaron in Einstein frame as
\begin{eqnarray}
V(\phi) = \frac{3}{4}\,M_p^2\, M^2
\left( 1- e^{-\sqrt{\frac23}\,\frac{\phi}{M_p}} \right)^2 .
\label{S-pot}
\end{eqnarray}
This is the standard Starobinsky scalaron potential in the Einstein frame and we will employ this potential in the analysis of the warm inflation scenario throughout this work. 

\subsection{Cosmological equations in warm inflation scenario}
Having used the HS action (\ref{action-E}) in the Einstein frame with the flat FRW line element, the Friedmann equation of the warm inflation is written by,
\begin{eqnarray}
H^2 = \frac{1}{3\,M_p^2}\left( \frac{1}{2}\,\dot\phi^2 + V(\phi) + \rho_r\right).
\end{eqnarray}
The Klein-Gordon equation of the scalaron field, $\phi$ with the dissipative term, $\Gamma$ due to the warm inflation scenario is governed by
\begin{eqnarray}
\ddot\phi + 3H\,\dot\phi + V' = -\Gamma\,\dot\phi\,.
\end{eqnarray}
While the conservation of the radiation matter is read
\begin{eqnarray}
\dot\rho_r + 4H\,\rho_r = \Gamma\,\dot\phi^2\,.
\end{eqnarray}
According to the finite temperature field theory analysis in the supersymmetry models, one obtains the general form of the dissipative parameter as \cite{Berera:1998gx,Berera:2001gs,Zhang:2009ge,BasteroGil:2011xd}
\begin{eqnarray}
\Gamma = C_m\,\frac{T^m}{\phi^{m-1}}\,.
\label{general-gamma}
\end{eqnarray}
The dissipative parameter, $\Gamma$ responds to the friction of the inflaton field in the thermal bath in the warm inflationary universe. In addition, the $C_m$ is the constant encoding the inflaton's microscopic effect of the dissitive dynamics and the $m$ is the integer number. In particular, the high temperature supersymmetric model is governed by $m=1$ whereas $m=3$ is responded to the low temperature supersymmetric model \cite{Zhang:2009ge}. 
In the following subsections. We will consider the dissipative parameter with the slow-roll approximation framework for $m=1$ which corresponds to a so-called warm little inflation.

The warm inflationary universe in the slow-roll regime, we can re-write the Firedmann equation as well as the equations of motion for the scalaron (inflaton) and the radiation matter as
\begin{eqnarray}
H^2 &\approx& \frac{1}{3 M_p^2}\,V(\phi)\,,
\label{SR-friedmann}
\\
\dot\phi &\approx& -\frac{V'(\phi)}{3H(1+Q)}\,,\qquad Q\equiv \frac{\Gamma}{3H}\,,
\label{SR-KG}
\\
\rho_r &\approx& \frac{\Gamma\,\dot\phi}{4H}\,,\qquad \rho_r = C_r\,T^4\,,
\label{SR-rad}
\end{eqnarray}
where the $Q$ is called a dissipative coefficient and $C_r = g_*\,\pi^2/30$. To obtain above equations, the following approximations have been used
\begin{eqnarray}
\rho_r &\ll& \rho_\phi\,,\qquad 
\rho_\phi = \frac12\,\dot\phi^2 + V\,,
\\
\dot\phi^2 &\ll& V(\phi)\,,
\\
\ddot\phi &\ll& 3H\left( 1 + Q\right)\dot\phi\,,
\\
\dot\rho_r &\ll& 4H\,\rho_r\,,
\end{eqnarray}
as usually done in the slow-roll scenario. It is more convenient to consider the warm inflation into two regimes as
\begin{eqnarray}
Q &\gg& 1\,,\qquad {\rm strong~regime}\,,
\\
Q &\ll& 1\,,\qquad {\rm weak~regime}\,.
\end{eqnarray}
More importantly, we can re-write the temperature in terms of the scalaron field, $\phi$ by using the Eqs. (\ref{general-gamma},\ref{SR-friedmann},\ref{SR-KG},\ref{SR-rad}) in the general $m$ integer values. One finds
\begin{eqnarray}
T &=& \left( \frac{V'^{\,2}\,\phi^{m-1}}{4H\,C_m\,C_r}\right)^{\frac{1}{4+m}}\,,\qquad {\rm for}\,~Q\gg 1\,,
\label{T-phi-strong}
\\
T &=& \left( \frac{C_m\,V'^{\,2}\,\phi^{1-m}}{36H^3\,C_r}\right)^{\frac{1}{4-m}}\,,\qquad {\rm for}\,~Q\ll 1\,.
\label{T-phi-weak}
\end{eqnarray}
Next, we provide the slow-roll parameters in the warm inflation for general $m$ and they read,
\begin{eqnarray}
\varepsilon &=& \frac{M_p^2}{2}\left( \frac{V'}{V}\right)^2\,,\quad \eta = M_p^2\,\frac{V''}{V}\,,\quad \beta = M_p^2\left( \frac{V'\,\Gamma'}{V\,\Gamma}\right)\,.
\label{SR-parameters}
\end{eqnarray}
The inflationary phase of the universe occurs under the following conditions
\begin{eqnarray}
\varphi \ll 1 + Q\,,\qquad \eta \ll 1 + Q\,,\qquad \beta \ll 1 + Q\,.
\end{eqnarray}
Moreover, the e-folding number, $N$ can be written in two regimes as
\begin{eqnarray}
N = \int_{\phi_{\rm end}}^{\phi_N}\frac{Q\,V}{V'}\,d\phi\,,\quad {\rm for}\quad Q\gg 1\,,\quad {\rm and}\quad N = \int_{\phi_{\rm end}}^{\phi_N}\frac{V}{V'}\,d\phi\,,\quad {\rm for}\quad Q\ll 1\,.
\end{eqnarray}
The power spectrum of the warm inflation has been calculated by Refs. \cite{Hall:2003zp,Ramos:2013nsa,BasteroGil:2009ec,Taylor:2000ze,DeOliveira:2001he,Visinelli:2016rhn}
and it reads,
\begin{eqnarray}
\Delta_{\mathcal{R}} = \left( \frac{H_N^2}{2\pi\dot\phi_N}\right)^2\left( 1 + 2n_N +\left(\frac{T_N}{H_N}\right)\frac{2\sqrt{3}\,\pi\,Q_N}{\sqrt{3+4\pi\,Q_N}}\right),
\end{eqnarray}
where the subscript $``N"$ is labeled for all quantities estimated at the Hubble horizon crossing and $n = 1/\big( \exp{H/T} - 1 \big)$ is the Bose-Einstein distribution function. In addition, the scalar spectral index is defined by
\begin{eqnarray}
n_s - 1 = \frac{d\ln \Delta_{\mathcal{R}}}{d\ln k}\Bigg|_{k=k_N} = \frac{d\ln \Delta_{\mathcal{R}}}{dN}\,,
\end{eqnarray}
with $\ln k\equiv a\,H =N$\,. The tensor-to-scalar ratio of the pertubation, $r$ can be calculated via the following formular
\begin{eqnarray}
r = \frac{\Delta_T}{\Delta_{\mathcal{R}}}\,,
\end{eqnarray}
where $\Delta_T$ is the power spectrum of the tensor perturbation and it takes the same form as the standard (cold) inflation picture, i.e. $\Delta_T = 2H^2/\pi^2M_p^2$. In the warm inflationary universe, the $r$ parameter has been determined in terms of the slow-roll parameters in both strong and weak regimes for $T\gg H$ limit by Refs \cite{Hall:2003zp,Ramos:2013nsa} as
\begin{eqnarray}
r = \frac{32\,\varepsilon}{\sqrt{3\,\pi}\,Q^{5/2}}\,,\quad {\rm for}~Q\gg 1\,,\qquad {\rm and} \qquad
r = 16\,\varepsilon\,,\quad {\rm for}~Q\ll 1\,.
\label{tensor-scalar-ratio}
\end{eqnarray}
In this work, we consider up to the first order of the $Q$ correction for $r$ parameter. As results, we note that only the strong regime is corrected by $Q$ whereas the the dissipative coefficient $Q$ does not play the role in the weak regime. Moreover, the $n_s$ is evaluated in the simple analytical forms for strong and weak regimes by Refs \cite{Hall:2003zp,Ramos:2013nsa}. Up to first order correction of the $Q$ parameter, the $n_s$ are given by
\begin{eqnarray}
n_s &=& 1 -\frac{1}{Q}\left( \frac94\,\varepsilon - \frac32\,\eta + \frac94\,\beta\right), \quad\qquad Q\gg1\,,
\\
n_s &=& 1 -6\,\varepsilon + 2\,\eta + \frac{1}{Q}\left( 8\,\varepsilon - 2\,\beta\right), \qquad Q\ll1\,.
\label{ns-form}
\end{eqnarray}
Nex, we will compute all relevant inflationary observables by considering $m=1$ dissipative parameter model in both strong and weak dissipative regimes. The dissipative parameter for $m=1$ model reads
\begin{eqnarray}
\Gamma = C_1\,T\,.
\label{model-Gamma}
\end{eqnarray}
As mentioned earlier, this model is related to the high temperature in supersymmetric models and also known as warm little inflation \cite{Bastero-Gil:2016qru}. More interestingly, the inflaton in this scenario coresponds to the pseudo-Goldstone boson from the broken symmetry and it is analogy to the little Higgs mechanism in the electroweak symmetry breaking framework. 

The slow-roll parameters of the $\Gamma = C_1\,T$ model are given by,
\begin{eqnarray}
\varepsilon &=& \frac{4}{3\left[ e^{\sqrt{\frac23}\,\frac{\phi}{M_p}} - 1\right]^2}\,,\qquad \eta = \frac43\,\frac{\left[2 - e^{\sqrt{\frac23\,\frac{\phi}{M_p}}}\right]}{\left[ e^{\sqrt{\frac23}\,\frac{\phi}{M_p}} - 1\right]^2}\,,
\nonumber\\
\beta &=& \frac{4}{15}\,\frac{\left[2 - 3\,e^{\sqrt{\frac23}\,\frac{\phi}{M_p}}\right]}{\left[ e^{\sqrt{\frac23}\,\frac{\phi}{M_p}} - 1\right]^2}\,,\quad {\rm for}\quad Q\gg 1\,,
\nonumber\\
\beta &=& \frac{4}{9}\,\frac{\left[1 - 2\,e^{\sqrt{\frac23}\,\frac{\phi}{M_p}}\right]}{\left[ e^{\sqrt{\frac23}\,\frac{\phi}{M_p}} - 1\right]^2}\,,\quad \;\;{\rm for}\quad Q\ll 1\,.
\label{slow-roll-model}
\end{eqnarray}
Before we proceed the theoretical results to be confronted with the data. It is worth estimating $Q$ in both strong and weak limits. By using Eqs.(\ref{SR-KG}), (\ref{T-phi-strong}) and (\ref{model-Gamma}). we find the $Q$ for the strong limit as,
\begin{eqnarray}
Q = \left[\frac{2^3}{3^4}\,\frac{\Theta\,e^{2\chi}}{\left(e^{\chi} - 1 \right)^4} \right]^{\frac15}\,,\qquad {\rm for}\quad Q\gg 1\,,
\label{Q-strong}
\end{eqnarray}
where we have defined a new parameter $\Theta \equiv \left( {C_1^4\,M_p^2}/{C_r\,M^2}\right)$ and  $\chi \equiv \sqrt{2/3}\,\phi/M_p$\,. On the other hand, the $Q$ in the weak regime, $Q\ll 1$ can be found by using Eqs. (\ref{SR-KG}), (\ref{T-phi-weak}) and (\ref{model-Gamma}). It is given by
\begin{eqnarray}
Q = \frac23\left[\frac{\Theta\,e^{2\chi}}{3\left(e^{\chi} - 1 \right)^4} \right]^{\frac13}\,,\qquad {\rm for}\quad Q\ll 1\,.
\label{Q-weak}
\end{eqnarray}

The warm inflation will be stop when the following conditions are satisfied,
\begin{eqnarray}
\varepsilon = 1 + Q\,,\qquad \eta = 1 + Q\,,\qquad \beta = 1 + Q\,.\label{sr}
\end{eqnarray}
In the latter will consider the end of the warm inflation for two cases, i.e., strong $Q\gg 1$ and weak $Q\ll 1$ approximation. We start with the strong regime. At the end of inflation, one finds from Eq.(\ref{sr})
\begin{eqnarray}
\varepsilon_{\rm end} \approx Q_{\rm end} \quad \Longrightarrow\quad  \frac{4}{3\left( e^{\chi} - 1\right)^2} \approx \left[\frac{2^3}{3^4}\,\frac{\Theta\,e^{2\chi}}{\left(e^{\chi} - 1 \right)^4} \right]^{\frac15}\,.
\end{eqnarray}
From the above equality we can solve to obtain the value of the inflaton field (scalaron) at the end of inflation to yield 
\begin{eqnarray}
\phi_{\rm end} \approx \sqrt{\frac32}\frac{M_p}{8}\ln\left( \frac{2}{3}\,\frac{4^3}{\Theta}\right),
\end{eqnarray}
where the large field approximation has been done via $e^\chi\pm1 \approx e^\chi$ with $\chi = \sqrt{2/3}\,\phi/M_p$\,. Moreover, the inflaton field at the Hubble horizon crossing in the strong regime, $\phi_{N}$, can be determined to obtain
\begin{eqnarray}
N &=& \frac{1}{M_p^2}\int_{\phi_{\rm end}}^{\phi_N}\frac{Q\,V}{V'}\,d\phi
\nonumber\\
&\approx& \frac52\left( \frac{C_1^4\,M_p^2}{18^2\,C_r\,M^2}\right)^{\frac15}\left( e^{\frac35\sqrt{\frac23}\,\frac{\phi_N}{M_p}} -e^{\frac35\sqrt{\frac23}\,\frac{\phi_{\rm end}}{M_p}} \right)
\nonumber\\
&\approx& \frac52\,\left(\frac{\Theta}{18^2}\right)^{\frac15}\,e^{\frac35\sqrt{\frac23}\,\frac{\phi_N}{M_p}}=\frac54\,\bigg[ \frac{2^3}{3^4}\,\Theta\bigg(\sqrt{\frac{4}{3\,\varepsilon}}\,\bigg)^{3}\,\Bigg]^{\frac15}\,,
\label{N-strong}
\end{eqnarray}
where $\Theta \equiv \left( {C_1^4\,M_p^2}/{C_r\,M^2}\right)$ and the condition $\phi_N\gg\phi_{\rm end}$ has been applied. This leads to
\begin{eqnarray}
e^{\chi_N} = \frac{12}{5}\left( \frac{6}{5^2}\,\frac{N^5}{\Theta}\right)^{\frac13}\,,\quad \Longrightarrow \quad \phi_N = \sqrt{\frac32}\,\frac{M_p}{3}\,\ln\left(8\frac{6^4}{5^5}\, \frac{N^5}{\Theta}\right)\,,
\label{expphi-N-strong}
\end{eqnarray}
where $\chi_N\equiv \sqrt{2/3}\,{\phi_N}/{M_p}$\,. As done above, we therefore can re-write the slow-roll parameters in terms of the {\it e}-folding number, $N$, by using the large field approximation in the strong $Q$ limit via
\begin{eqnarray}
\varepsilon \approx \frac{5^3}{4\cdot\,3^3}\left( \frac{5^2}{6}\,\frac{\Theta}{N^5}\right)^{\frac23},
\quad \eta \approx - \sqrt{\frac{4\,\varepsilon}{3}} = -\frac{5\sqrt{5}}{9}\left( \frac{5^2}{6}\,\frac{\Theta}{N^5}\right)^{\frac13},\quad \beta \approx -\frac{2\,\sqrt{3\,\varepsilon}}{5} = -\frac{\sqrt{5}}{3}\left( \frac{5^2}{6}\,\frac{\Theta}{N^5}\right)^{\frac13}.
\label{SR-strong-N}
\end{eqnarray}

On the other hand, in the weak regime $Q\ll1$, the end of inflation yields
\begin{eqnarray}
\varepsilon_{\rm end} &\approx& 1\,\quad \Longrightarrow \quad \phi_{\rm end} \approx 0.18\,M_p\,. 
\end{eqnarray}
While the e-folding number in the weak regime is given by
\begin{eqnarray}
N &=& \frac{1}{M_p^2}\int_{\phi_{\rm end}}^{\phi_N}\frac{V}{V'}\,d\phi
\nonumber\\
&\approx& \frac34\,\left( e^{\sqrt{\frac23}\,\frac{\phi_N}{M_p}} -e^{\sqrt{\frac23}\,\frac{\phi_{\rm end}}{M_p}} \right)
\nonumber\\
&\approx& \frac34\,e^{\sqrt{\frac23}\,\frac{\phi_N}{M_p}}=\frac{3}{4}\sqrt{\frac{4}{3\,\varepsilon}}\,.
\label{N-weak}
\end{eqnarray}
where approximations $e^\chi\pm1 \approx e^\chi$ and $\phi_N\gg\phi_{\rm end}$ are once assumed. As a result, we find 
\begin{eqnarray}
\phi_N = \sqrt{\frac32}\,M_p\,\ln\left( \frac43\,N\right).
\label{phi-N-weak}
\end{eqnarray}
In addition, we also re-write the slow-roll parameters in terms of $N$ in the weak regime, $Q\ll 1$ as
\begin{eqnarray}
\varepsilon \approx \frac{3}{4\,N^2}\,,\qquad \eta \approx - \sqrt{\frac{4\,\varepsilon}{3}} = -\frac{1}{N}\,,\qquad \beta \approx -\frac23\,\sqrt{\frac{4\,\varepsilon}{3}} =-\frac{2}{3\,N}\,.
\label{SR-weak-N}
\end{eqnarray}


\section{Confrontation with the Planck 2018 data}
\label{compare-data}
In order to confront with the observational data, we need to compute the relevant observables i.e., the tensor to scalar perturbation raito, $r$ and the spectral index, $n_s$ by using Eqs. (\ref{tensor-scalar-ratio}) and (\ref{ns-form}), respectively. Moreover, we separate our investigations into two cases for the strong ($Q\gg 1$) and weak ($Q\ll 1$) limits in the latter. 

We will constrain our scalaron potential with the COBE normalization condition [cite] for fixing
parameters in the Higgs-Starobinsky model.  To generate the observed amplitude of the density perturbation ($A_{s}$), the potential must satisfy the COBE renormalization at horizon crossing $\phi=\phi_{N}$:
\begin{eqnarray}
\frac{V}{\varepsilon}\bigg|_{\phi=\phi_{N}} \simeq (0.0276\,M_{p})^{4}=\frac{3{\cal M}}{4}e^{2\chi_N}\,,
\label{1r-strong}
\end{eqnarray}
where we have defined ${\cal M}\equiv 3M^{2}_{p}M^{2}/4,\,\chi_N\equiv \sqrt{2/3}\,\phi_N/M_p$ and this is used to impose a constraint on the mass scale $M$ given in Eq.(\ref{Mde}). 
\begin{figure}[!h]	
	\includegraphics[width=12cm]{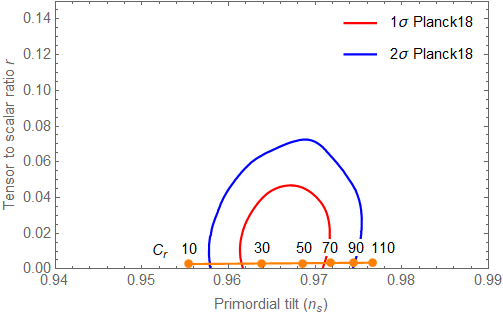}
	\centering
	\caption{We compare the theoretical predictions  of the strong limit $Q>1$ given in Eqs.(\ref{rst1}) and (\ref{nsst1}) in the ($r-n_{s}$) plane for different values
of $C_{r}$ using $C_{1}=6.0\times 10^{-1}$ and $N=60$ with Planck’18 results for TT, TE, EE, +lowE+lensing+BK15+BAO.}
	\label{stplot}
\end{figure}
\begin{figure}[!h]	
	\includegraphics[width=12cm]{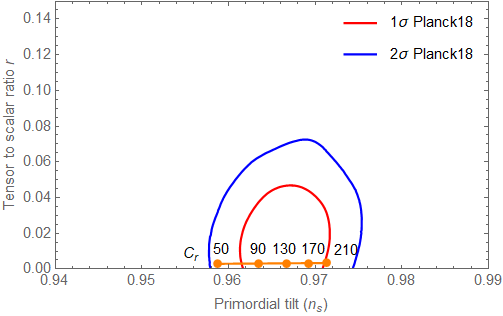}
	\centering
	\caption{We compare the theoretical predictions  of the strong limit $Q>1$ given in Eqs.(\ref{rst1}) and (\ref{nsst1}) in the ($r-n_{s}$) plane for different values
of $C_{r}$ using $C_{1}=8.0\times 10^{-1}$ and $N=60$ with Planck’18 results for TT, TE, EE, +lowE+lensing+BK15+BAO.}
	\label{stplot1}
\end{figure}
\subsection{Strong regime, $Q\gg 1$}
In the strong regime of the coefficient $Q$ given in Eq.(\ref{Q-strong}), we calculate the tensor-to-scalar ration by using the definition of $r$ in Eq.(\ref{tensor-scalar-ratio}) and the values of the slow-roll parameters given in Eq.(\ref{slow-roll-model}), The $r$ parameter reads
\begin{eqnarray}
r = \frac{32}{\sqrt{3\,\pi}}\left[\frac{4}{3\left( e^{\chi_N} - 1\right)^2} \right]\left[\frac{2^3}{3^4}\,\frac{\Theta\,e^{2\chi_N}}{\left(e^{\chi_N} - 1 \right)^4} \right]^{-\frac12} \approx \frac{32}{\sqrt{3\,\pi}}\,\frac{6\,e^{-\chi_N}}{\sqrt{2\,\Theta}}\,,
\label{r-strong}
\end{eqnarray}
while the spectral index for $Q\gg 1$ limit is given by
\begin{eqnarray}
n_s = 1 - \left[\frac{2^3}{3^4}\,\frac{\Theta\,e^{2\chi_N}}{\left(e^{\chi_N} - 1 \right)^4} \right]^{-\frac15}\left[ \frac{3}{\left( e^{\chi_N} - 1\right)^2} -\frac{2\left(2 - e^{\chi_N}\right)}{\left( e^{\chi_N} - 1\right)^2} +\frac35\,\frac{\left(2 - 3\,e^{\chi_N}\right)}{\left( e^{\chi_N} - 1\right)^2}\right].
\end{eqnarray}
It is more convenient to express $r$ and $n_{s}$ in terms of a number of {\it e}-folds $N$ by using Eqs.(\ref{N-strong}) and (\ref{SR-strong-N}). For $Q\gg 1$ case, they become
\begin{eqnarray}
r &=& \frac{40}{\sqrt{\pi}}\left( \sqrt{\frac{2}{3^5\,\Theta}}\,\frac{5^2}{N^5}\right)^{\frac13}\,,\label{rst}\\
n_s &=& 1 - \frac{19}{4\,N}+\frac{5^2}{4^2}\left( \frac{5^2}{6}\,\frac{\Theta}{N^8}\right)^{1/3}.\label{nsst}
\end{eqnarray}
Additionally, substituting $\phi_N$ from Eq.(\ref{expphi-N-strong}) into Eq.(\ref{1r-strong}), we obtain
\begin{eqnarray}
M^{2}\approx \frac{1.12162\times 10^{-4}\,C^{8/5}_{1} M_{p}^2}{C_{r}^{2/5} N^2}. \label{M2ss}
\end{eqnarray}
Using the observational constraint of $r<0.01$, we find from Eq.(\ref{rst}) that
\begin{eqnarray}
C_{r}<7.30\times 10^{-29}C_{1}^4 N^{20}.\label{cr}
\end{eqnarray}
For example, we assume $C_{1}=4.0\times 10^{-2}$ and $N=60$. Thus in order to satisfy Eq.(\ref{cr}), we obtain $C_{r}<68.3$. We instead write Eq.(\ref{rst}) and Eq.(\ref{nsst}) in terms of $N,\,C_{r}$ and $C_{1}$ to obtain
\begin{eqnarray}
r &=& 6.51\,\bigg(\frac{C_{r}^{3/5}}{C_{1}^{12/5}N^{12}}\bigg)^{1/6}\,,\label{rst1}\\
n_s &=& 1 - \frac{19}{4\,N}+ 52.14\, \bigg(\frac{C_{1}^{12/5}}{C_{r}^{3/5}N^{6}}\bigg)^{1/3}.\label{nsst1}
\end{eqnarray}

We compare our predictions given by Eq.(\ref{rst1}) and Eq.(\ref{nsst1}) for different values of $C_{r}$ with Planck’18 results for TT, TE, EE, +lowE+lensing+BK15+BAO. As an example, we consider $C_{1}=6.0\times 10^{-1},\,8.0\times 10^{-1}$ and keep a number of {\it e}-folds fixed at $N=60$. We use $C_{1}=6.0\times 10^{-1}$ in Fig.(\ref{stplot}). We find that our results show very small values of $r$. For example, we find $r=3.28\times 10^{-3}$ for $N=60,\,C_{1}=6.0\times 10^{-1}$ and $C_{r}=50$. In order to have
the predictions fit well inside the $1\,\sigma$ regions of the Planck 2018 data, values of $C_r$ are constrained between $24 < C_r < 64$ using $N=60,\,C_{1}=6.0\times 10^{-1}$. 

Additionally, we consider another value of $C_{1}=8.0\times 10^{-1}$. Our results are displayed in Fig.(\ref{stplot1}). In this case, we have $r=3.10\times 10^{-3}$ for $N=60,\,C_{1}=8.0\times 10^{-1}$ and $C_{r}=90$. In order to have
the predictions agree well with the $1\,\sigma$ regions of the Planck 2018 data, values of $C_r$ are constrained between $75 < C_r < 204$ using $N=60,\,C_{1}=8.0\times 10^{-1}$. We discover for the strong limit that the thermal bath makes significant effects to the inflationary observables $r$ and $n_{s}$ and our results are different from those found in the existing literature \cite{Bastero-Gil:2016qru,Bastero-Gil:2018uep,Benetti:2016jhf,Bastero-Gil:2017wwl,Arya:2018sgw}. 

Moreover, for the strong limit, we can further use the scalaron mass parameter to constrain underlying parameters $\alpha,\,\lambda,\,\xi$ by using the relation:
\begin{eqnarray}
M^2 = \frac{M_p^2}{12\left(\alpha + 3\,\xi^2/(2\,\lambda)\right)}\,. \label{M2s}
\end{eqnarray}
Using Eq.(\ref{M2ss}), we find
\begin{eqnarray}
\frac{M_p^2}{12\left(\alpha + 3\,\xi^2/(2\,\lambda)\right)}\approx \frac{1.12162\times 10^{-4}\,C^{8/5}_{1} M_{p}^2}{C_{r}^{2/5} N^2}, \label{M2ss1}
\end{eqnarray}
which yields
\begin{eqnarray}
\lambda=-\frac{1.5\, C_{1}^{8/5}\, \xi ^2}{ C_{1}^{8/5}\,\alpha - 742.97\, C_{r}^{2/5} N^2}.
\label{lambda-warm-inflation}
\end{eqnarray}
Notice that the thermal bath effects are present in the above relation and also play significant role to the values of $\alpha,\,\lambda,\,\xi$. We find for example using $C_{r}=90,\,C_{1}=8.0\times 10^{-1}$ with $N=60,\,\xi=10,000$ and this leads to
\begin{eqnarray}
\lambda \sim -\frac{1.05\times 10^8}{0.70\,\alpha -1.62\times 10^7}  \quad \Longrightarrow \quad
\alpha_{\rm warm} \sim -1.50\times 10^{10} \quad {\rm for}\quad \lambda = 0.01\,.
\end{eqnarray}

Interestingly, in order to be satisfied with the Planck data, our results in the strong regime $Q\gg 1$ exceed the upper bound of the $C_1\lesssim 0.02$ in the original model of the warm little inflation \cite{Bastero-Gil:2016qru,Bastero-Gil:2018uep}. However, the upper bound of the $C_1$ parameter can be relaxed in order to obtain results compatible with the observational data in the linear temperature dissipative model. For instance, this is found in non-minimal coupling Higgs warm inflation model \cite{Kamali:2018ylz} and the power-law plateau warm inflation potential \cite{Jawad:2017rkq}. 
Having compared with the constraint of the HS model in cold inflation scenario $\alpha_{\rm cold}$ given by \cite{Samart:2018mow}
\begin{eqnarray}
\alpha_{\rm cold} &\sim& -\,1.45\times 10^{10}\,,\quad {\rm for} \quad \lambda = 0.01\,,
\end{eqnarray}
we find that the values of $\alpha$ of the $R^2$ term constrained from warm inflation is bigger than those of $\alpha$ obtained from cold inflation about 3.5\%\,. In addition, we still obtain a large value of the $\alpha$ parameter which is required from the density perturbation for successful inflation \cite{Netto:2015cba}.
\begin{figure}[!h]	
	\includegraphics[width=12cm]{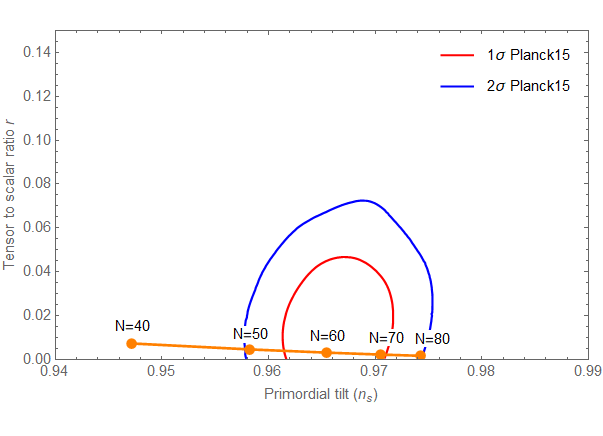}
	\centering
	\caption{We compare the theoretical predictions  of the weak limit $Q<1$ given in Eqs.(\ref{wr}) and (\ref{wns}) in the ($r-n_{s}$) plane for different values of $N$ using $C_{r}=70$ and $C_{1}=2.0\times 10^{-7}$ with Planck’18 results for TT, TE, EE, +lowE+lensing+BK15+BAO.}
	\label{stplot2}
\end{figure}
\subsection{Weak regime, $Q\ll 1$}
The tensor to scalar perturbation ration in the weak regime $Q\ll 1$ is taken in to the following form, 
\begin{eqnarray}
r = 16\left[\frac{4}{3\left( e^{\chi_N} - 1\right)^2} \right].
\end{eqnarray}
According to the above equation, the $r$ parameter has the same for as the standard (cold) inflation result. For the spectral index, $n_s$ in the weak limit is written by,
\begin{eqnarray}
n_s &=& 1 - \frac{24}{3\left( e^{\chi_N} - 1\right)^2} + \frac83\,\frac{\left(2 - e^{\chi_N}\right)}{\left( e^{\chi_N} - 1\right)^2} 
\nonumber\\
&+& \frac{2}{3}\left[\frac{\Theta\,e^{2\chi_N}}{3\left(e^{\chi_N} - 1 \right)^4} \right]^{\frac13}\left[ \frac{32}{3\left( e^{\chi_N} - 1\right)^2} - \frac{8}{9}\,\frac{\left(1 - 2\,e^{\chi_N}\right)}{\left( e^{\chi_N} - 1\right)^2}\right].
\end{eqnarray}
Again, it is more convenient to write $r$ and $n_{s}$ in terms of a number of {\it e}-folds $N$. For $Q\ll 1$ case, with help of Eqs.(\ref{N-weak}) and (\ref{SR-weak-N}), they read
\begin{eqnarray}
r &\approx& \frac{12}{N^{2}}\,,\label{wr}\\
n_s &\approx& 1-\frac{2}{N}-\frac{9}{2^2} +\frac{2}{3}\bigg(\frac{3 \left(1.72 \times 10^6 C_{1}^4\right)}{2^4 C_{r}}\bigg)^{1/3}\left(\frac{6}{N^2}+\frac{3}{2 N}\right).\label{wns}
\end{eqnarray}
Here we wrote $n_{s}$ in terms of parameters $C_{1},\,C_{r}$ and $N$. Moreover, we solve Eq.(\ref{1r-strong}) to obtain
\begin{eqnarray}
M^{2}=\frac{5.80\times 10^{-7}\,M_{p}^{2}}{N^{2}}.
\end{eqnarray}
We compare our predictions given by Eq.(\ref{wr}) and Eq.(\ref{wns}) for different values of $N$ with Planck’18 results for TT, TE, EE, +lowE+lensing+BK15+BAO. As an example, we use typical values of $C_{r},\,C_{1}$ as given in Ref.\cite{Kamali:2018ylz}. In this weak limit, we also find that the results show very small values of $r$. In order to have
the predictions fit well inside the $1\,\sigma$ regions of the Planck 2018 data, values of $C_r$ are constrained between $55 < N < 70$ using $C_r=70,\,C_{1}=2.0\times 10^{-7}$. We discover for the weak limit that the thermal bath makes makes negligible effects to the inflationary observables $r$ and $n_{s}$ due to a very tiny values of $C_{1}$ required.

For the weak limit, we can also use the the scalaron mass parameter to constrain underlying parameters $\alpha,\,\lambda,\,\xi$ using the relation:
\begin{eqnarray}
M^2 = \frac{M_p^2}{12\left(\alpha + 3\,\xi^2/(2\,\lambda)\right)}\,. \label{M2s}
\end{eqnarray}
Using Eq.(\ref{M2ss}), we find
\begin{eqnarray}
\frac{M_p^2}{12\left(\alpha + 3\,\xi^2/(2\,\lambda)\right)}\approx \frac{5.80\times 10^{-7}\,M_{p}^{2}}{N^{2}}, \label{M2sw1}
\end{eqnarray}
which yields
\begin{eqnarray}
\lambda=\frac{1.04\times 10^{-5} \xi ^2}{N^2-6.96\times 10^{-6} \alpha }.
\end{eqnarray}
We find for example using $ \lambda\approx \frac{1044}{3600-6.96\times 10^{-6} \alpha}$ with $N=60,\,\xi=10,000$. We find that values of underlying parameters $\alpha,\,\lambda,\,\xi$ are not affected by the thermal bath counterpart. 

\section{conclusion}\label{conclud}
In this work, we have demonstrated a class of
warm inflation scenario using HS gravity with a linear temperature of the dissipative parameter. We have studied the dynamics of the warm inflation in the Einstein frame and considered our analysis into two regimes, strong ($Q\gg 1$) and weak ($Q\ll 1$). We have calculated relevant observables in the warm inflation in order to compare to the observational data. In the strong regime, we have discovered that inflationary parameters $r$ and $n_{s}$ can be written in terms of the parameters $C_r$ and $C_{1}$ and hence they are affected by having the thermal bath, while in the weak regime the inflationary parameters are very weakly affected by the thermal bath. Therefore the thermal bath effects are approximately negligible in this regime. 

According to our analysis, we have found that the HS model in weak regime provides an excellent agreement with the data, whilst the thermal bath effects have played an significant role in the strong dissipative regime. The ranges of the parameters in HS model have been evaluated to make the predictions compatible with the Planck 2018 results. Consequently, we have also found that the Starobinsky gravitational coupling, $\alpha$ is slightly modified by the dissipative parameters $C_r$ and $C_{1}$ present in warm inflation. Interestingly, in order to be satisfied with the Planck data, our results of $C_1$ in the strong regime $Q\gg 1$ exceed the upper bound of $C_1\lesssim 0.02$ mentioned in the original model of warm little inflation \cite{Bastero-Gil:2016qru,Bastero-Gil:2018uep}. Finally, with the sizeable number of e-folds and proper choices of parameters, we have also discovered for the strong regime that the predictions of warm HS model present in this work are in very good agreement with the latest Planck 2018 results.

In addition, more models of the different/same dissipative parameter are interesting for future investigation. More importantly, further studies on the dynamics of the universe after radiation-dominated era might shed some light on the Hubble tension problem. We wish to address this topic for future study.  

\section*{Acknowledgements}
D. Samart is financially supported by National Astronomical Research Institute of Thailand (NARIT). P. Channuie acknowledged the Mid-Career Research Grant 2020 from National Research Council of Thailand under a contract No. NFS6400117.

\end{document}